\documentclass[11pt]{article}
\usepackage{authblk}
\usepackage{float}
\usepackage{graphicx} 
\usepackage{caption}
\usepackage{xcolor}
\usepackage{amsmath}
\usepackage{amssymb}
\usepackage{hyperref}
\usepackage{caption}
\hypersetup{
    colorlinks=true,
    linkcolor=blue,
    citecolor=blue,
    urlcolor=blue
}

\usepackage[a4paper,margin=1in]{geometry}
\usepackage{authblk}
\usepackage{hyperref}
\usepackage{titlesec}
\usepackage{setspace}
\usepackage{lmodern}

\titleformat{\section}{\large\bfseries}{\thesection.}{0.5em}{}

\title{\textbf{The Cultural Mapping and Pattern Analysis (CMAP) Visualization Toolkit:}\\
\textbf{Open Source Text Analysis for Qualitative and Computational Social Science}}

\author[1,2,3,4,5,6,7]{\textbf{Corey M. Abramson}\thanks{Equal contribution. ORCID: \href{https://orcid.org/0000-0001-6306-6910}{0000-0001-6306-6910}}}
\author[2,8]{\textbf{Yuhan (Victoria) Nian}\thanks{Equal contribution. Corresponding author. ORCID: \href{https://orcid.org/0009-0006-2603-7479}{0009-0006-2603-7479}}}

\affil[1]{Associate Professor of Sociology, Department of Sociology, Rice University, United States}
\affil[2]{Computational Ethnography Lab, Rice University, United States}
\affil[3]{Co-Director, Center for Computational Insights on Inequality and Society (CIISR), Rice University, United States}
\affil[4]{Affiliated Faculty, Institute of Health Resilience and Innovation (IHRI), Rice University, United States}
\affil[5]{Affiliated Faculty, Ken Kennedy Institute (Responsible AI and Scientific Computing), Rice University, United States}
\affil[6]{Faculty, Medical Cultures Lab, University of California San Francisco, United States}
\affil[7]{Affiliated Faculty, Center for Ethnographic Research, University of California Berkeley, United States}
\affil[8]{Department of Statistics, Rice University, United States}
\date{}
\begin{document}
\maketitle

\section*{Summary}

The \textbf{CMAP} (Cultural Mapping and Pattern Analysis) visualization toolkit is an open-source suite for analyzing and visualizing text data—from qualitative fieldnotes and in-depth interview transcripts to historical documents and web-scraped data such as message board posts or blogs. The toolkit is designed for scholars integrating pattern analysis, data visualization, and explanation in qualitative and/or computational social science (CSS).

Despite the existence of off-the-shelf commercial qualitative data analysis software, there remains a shortage of highly scalable open-source options capable of handling large datasets and supporting advanced statistical and language modeling.

The foundation of the toolkit is a pragmatic approach that aligns research tools with social science project goals—empirical explanation, theory-guided measurement, comparative design, or evidence-based recommendations—guided by the principle that research paradigms and questions should determine methods. Consequently, the CMAP visualization toolkit offers a wide range of possibilities through the adjustment of a relatively small number of parameters and allows seamless integration with other Python tools.

\section*{Statement of Need}

This software builds on sociological traditions of multi-method analysis, triangulation, and purposive computation to link levels of analysis and generate insights of scientific and practical importance \cite{dubois1899philadelphia, lamont2009workshop, small2011mixed}. Computational tools in this framework expand human inquiry, continuing a trajectory from statistical computing, qualitative data analysis software, CSS text analyses, and visualization to open science. The toolkit proceeds from the premise that computation is already embedded in research and daily life—from CAQDAS software to search algorithms—and can be used thoughtfully to advance sociological inquiry and ensure emergent technologies address pressing social issues \cite{abramson2025pragmatic, grimmer2022text, healy2014data, nelson2020computational, breiger2015scaling, dohan1998using, peponakis2023calculations, fourcade2024ordinal}.

CMAP includes cutting-edge visualization options that are open source and accessible to those without extensive Python programming experience, making it adaptable as both a pedagogical and research tool—addressing core issues of training and accessibility important for expanding CSS proficiencies for qualitative researchers \cite{abramson2025pragmatic}.

The toolkit supports advanced analytic methods appropriate for computational text analysis alongside in-depth readings—including co-occurrence, clustering, and embedding approaches—with visuals such as heatmaps, t-SNE dimensional reduction plots (analogous to scatter plots of words), semantic networks, word clouds, and more. Examples are compatible with common qualitative data sources and allow granular analysis that mirrors qualitative practices (at the level of words, sentences, and paragraphs) while scaling for large datasets produced by research teams.

CMAP visualizations are designed for integration into research papers and pedagogical applications, addressing the dearth of open-source software accessible to qualitative researchers seeking scalable analytical tools using established data visualizations with transparent statistical foundations. The toolkit runs efficiently on consumer-grade hardware without extensive setup, even when employing advanced features such as word embeddings.

The main paper outlines the organization and functions of the toolkit. Full mathematical details, related software resources, and representative scientific applications are provided in the Appendix.

\section*{CMAP Organization}

\noindent
CMAP can be run in either a Jupyter environment (via \href{https://github.com/Computational-Ethnography-Lab/cmap_visualization_toolkit.git}{GitHub}) or Google Colab (\href{https://colab.research.google.com/drive/1n90EDMSiXhIaOULUMPJ4W4hqdZCh1NQw?authuser=1#scrollTo=1jgH13I3xLbA&uniqifier=1}{Colab Link}). Colab is recommended for learning the methods and experimenting with public datasets. For sensitive data or extended development, users can clone the GitHub repository and run the included installation script locally:

\begin{verbatim}
git clone https://github.com/Computational-Ethnography-Lab/cmap_visualization_toolkit.git
cd cmap_visualization_toolkit
chmod +x install.sh
./install.sh
\end{verbatim}

\noindent
The repository contains several key files:

\begin{itemize}
  \item \texttt{.sh} --- installation and environment setup
  \item \texttt{.py} --- core mathematical functions for similarity, clustering, and network layout
  \item \texttt{.ipynb} --- the main program with workflows for importing, validating, cleaning, modeling, and visualizing text \cite{grimmer2022text, abramson2025pragmatic}
\end{itemize}

\begin{figure}[H]
    \centering
    \includegraphics[width=0.8\textwidth]{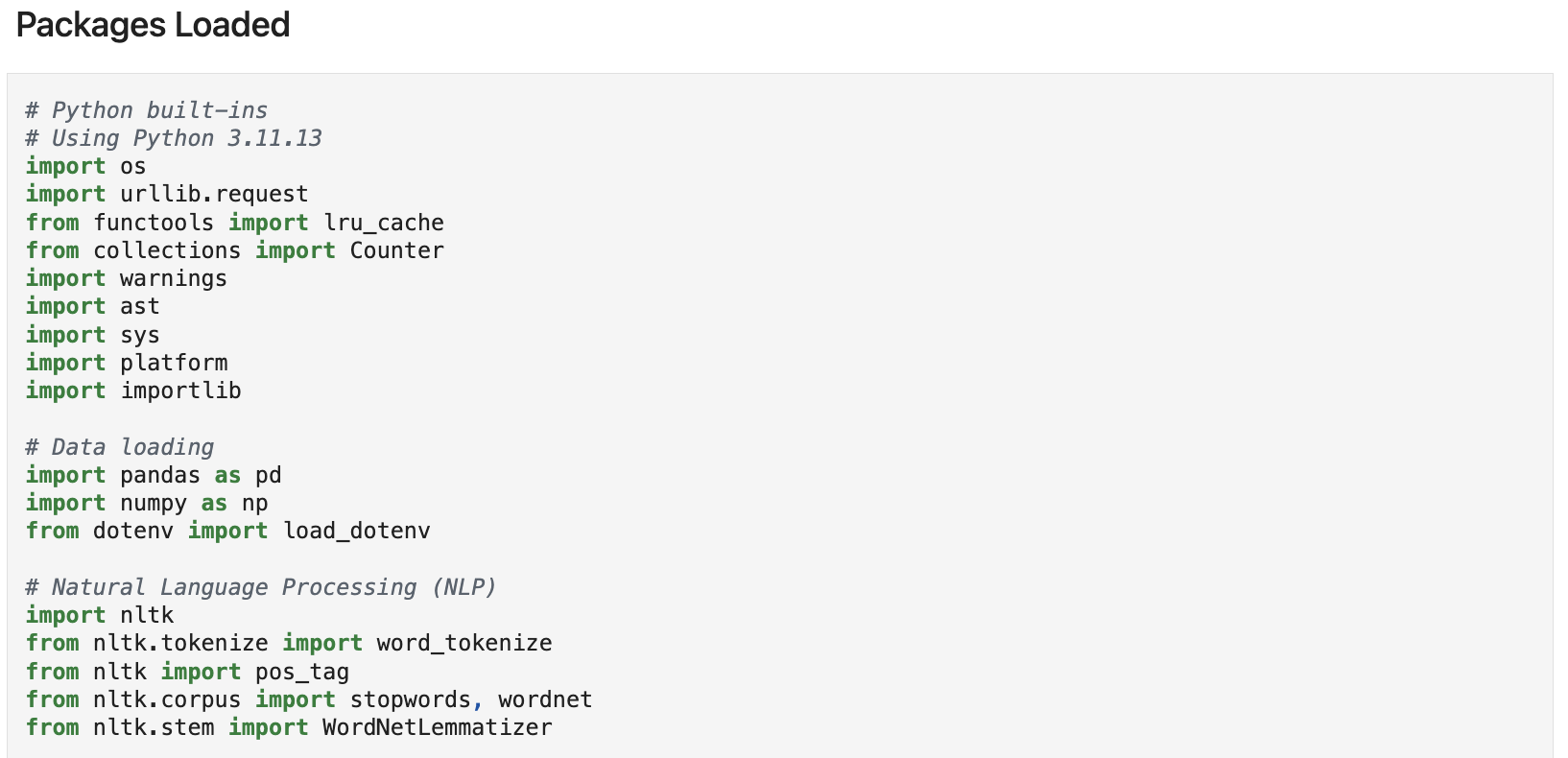}
    \caption{Package Imports.}
    \label{package_load}
\end{figure}

\begin{figure}[H]
    \centering
    \includegraphics[width=0.8\textwidth]{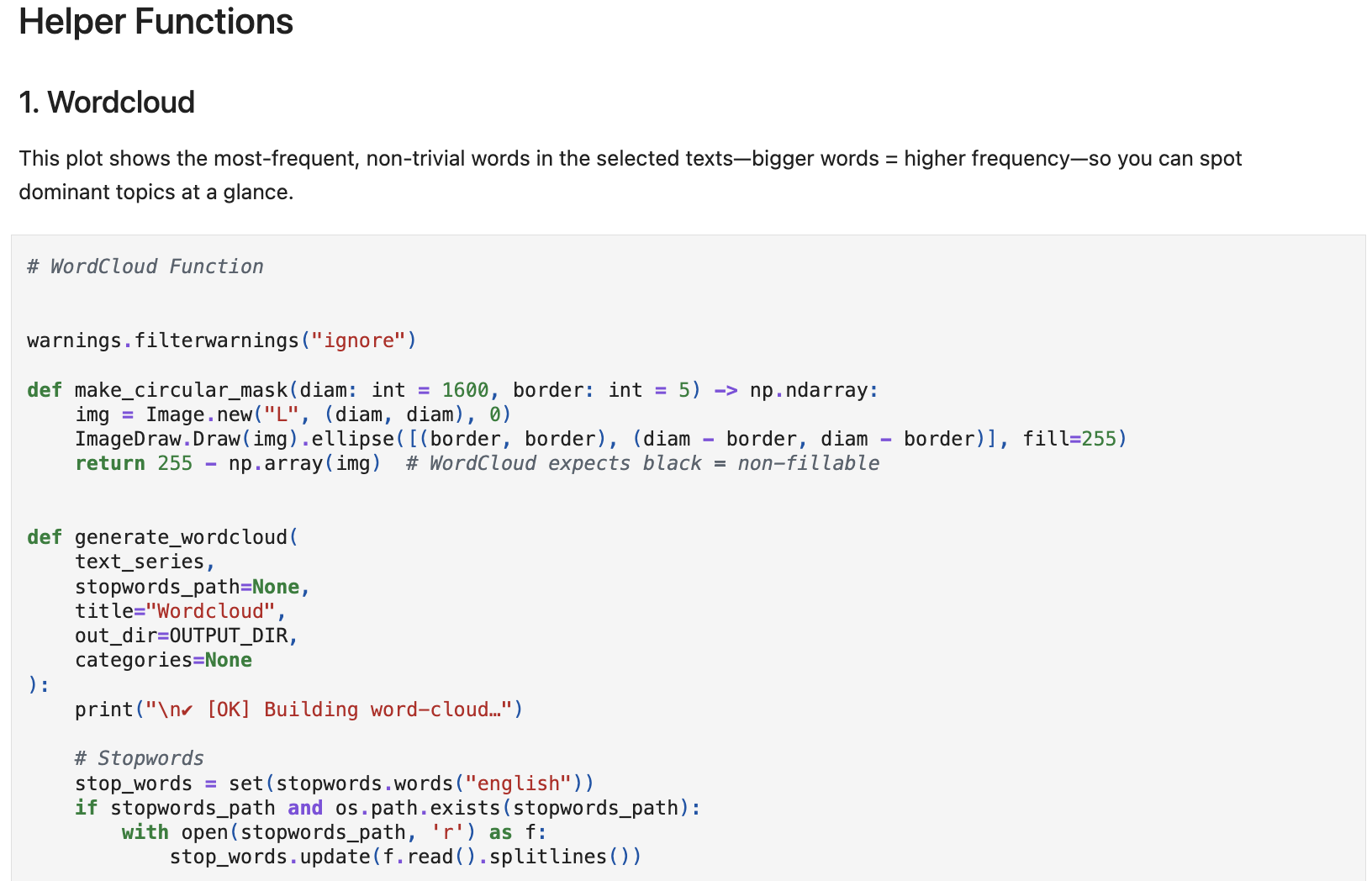}
    \caption{Helper Functions.}
    \label{helper_functions}
\end{figure}

\begin{figure}[H]
    \centering
    \includegraphics[width=0.8\textwidth]{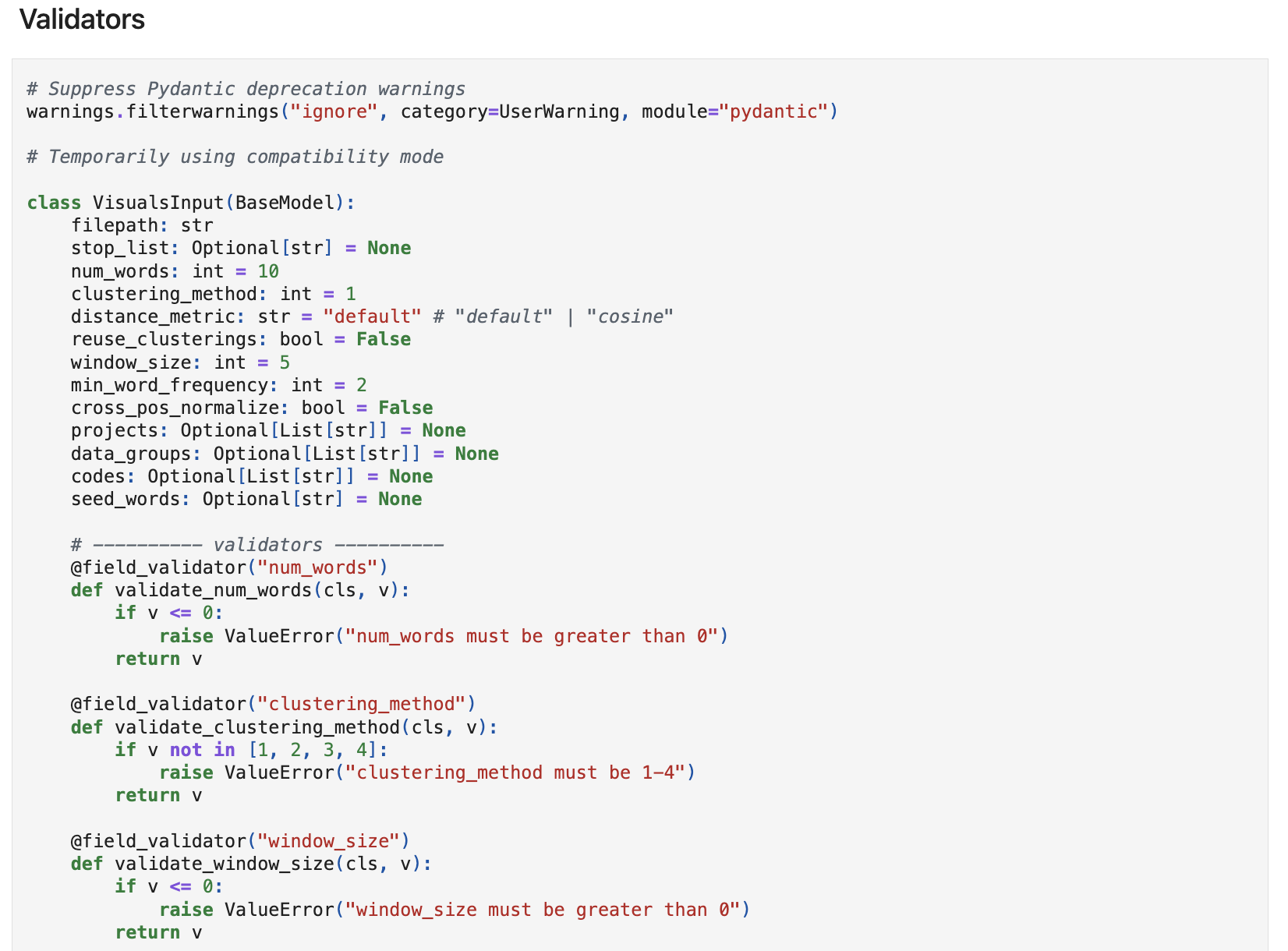}
    \caption{Validator.}
    \label{validator}
\end{figure}

\begin{figure}[H]
    \centering
    \includegraphics[width=0.8\textwidth]{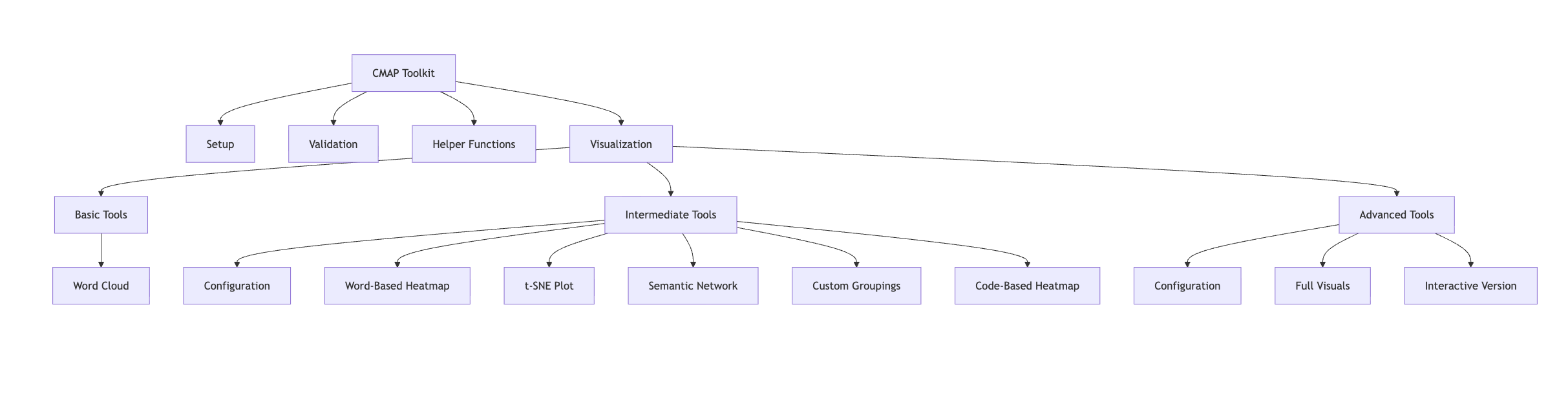}
    \caption{Program Organization of the CMAP Toolkit.}
    \label{flowchart}
\end{figure}

\noindent
As shown in \autoref{package_load}, \autoref{helper_functions}, and \autoref{validator}, the main program is organized into modular execution blocks (e.g., Imports, Validation, Helper Functions, Visualization), which correspond to each step in the text-visualization pipeline (\autoref{flowchart}). To illustrate this structure, we include example code screenshots from each section of the toolkit below. This modular design allows users to flexibly adapt CMAP for both small-scale classroom applications and large collaborative research projects.

\section*{Functions}

CMAP provides four options for measuring relationships between words or concepts, each emphasizing a different type of connection (for detailed mathematical implementations, see Appendix).

\begin{itemize}
    \item \textbf{RoBERTa (Semantic Similarity)}: Finds words used in conceptually similar ways using dynamic contextual embeddings \cite{liu2019roberta}. This method is best for uncovering analogies and latent meanings (e.g., \emph{success} $\rightarrow$ \emph{money}, \emph{happiness}, \emph{family}). The embedding model can be replaced with fine-tuned or specialized alternatives.
    
    \item \textbf{Co-occurrence (Jaccard or Cosine Similarity Distance)}: Best for identifying direct vocabulary associations within the same text segments (e.g., \emph{success} $\rightarrow$ \emph{hard work}, \emph{effort}).
    \begin{itemize}
        \item \textbf{Jaccard index}: Set-based, binary overlap score emphasizing whether words co-occur at all.
        \item \textbf{Cosine similarity}: Compares frequency-sensitive context vectors built from co-occurrence counts.
    \end{itemize}
    
    \item \textbf{PMI (Pointwise Mutual Information)}: Highlights words that co-occur more often than expected by chance. Best for identifying statistically significant pairings.
    
    \item \textbf{TF--IDF (Term Frequency--Inverse Document Frequency)}: Detects distinctive words that are unusually important within a given segment \cite{manning2008introduction, newman2010networks}. By default, CMAP applies cosine similarity to vector-based methods, balancing interpretability and sensitivity in accordance with common practices in computational social science. Alternative options (e.g., Jaccard overlap or raw-weighted TF--IDF) allow researchers to emphasize overlap, context, or frequency.
\end{itemize}

\section*{Visualization}

CMAP produces multiple visual outputs that allow researchers to explore relationships at different levels (words, sentences, paragraphs) and scale to large collaborative datasets. These visualizations mirror pragmatic mixed-methods principles while enabling scalable analysis and allowing users to adjust settings as needed.

\begin{figure}[H]
  \centering
  
  \begin{minipage}{0.5\textwidth}
    \centering
    \includegraphics[width=\linewidth]{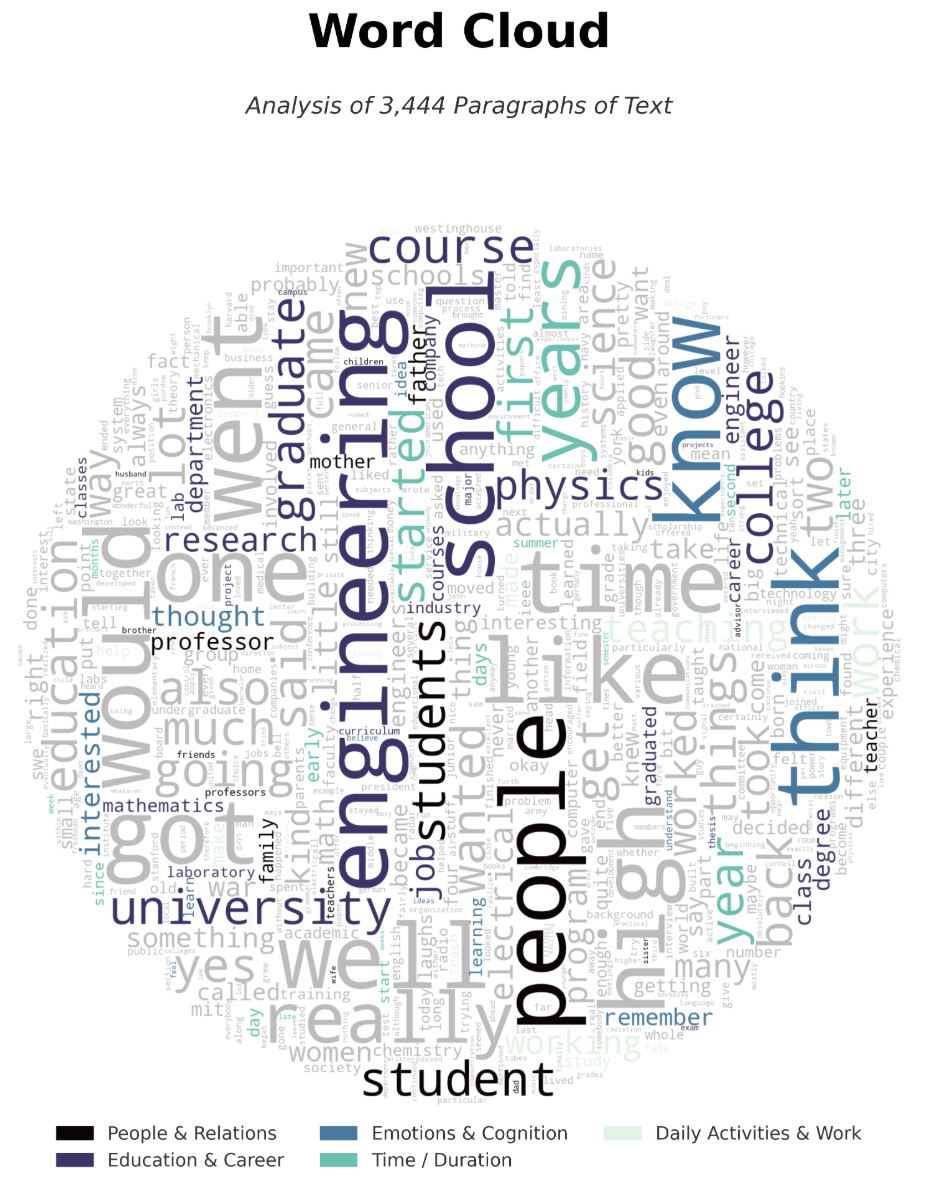}
    \captionof{figure}{Word Cloud}
    \label{word_cloud}
  \end{minipage}\hfill
  \begin{minipage}{0.5\textwidth}
    \centering
    \includegraphics[width=\linewidth]{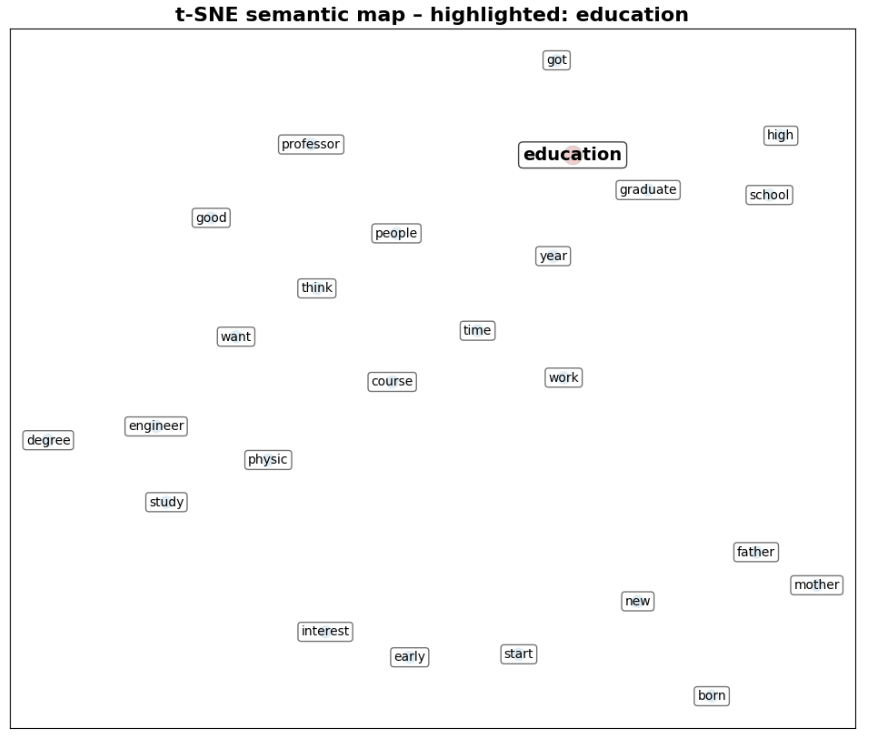}
    \captionof{figure}{t-SNE Plot}
    \label{tsne}
  \end{minipage}

  \vspace{1em}

  \begin{minipage}{0.5\textwidth}
    \centering
    \includegraphics[width=\linewidth]{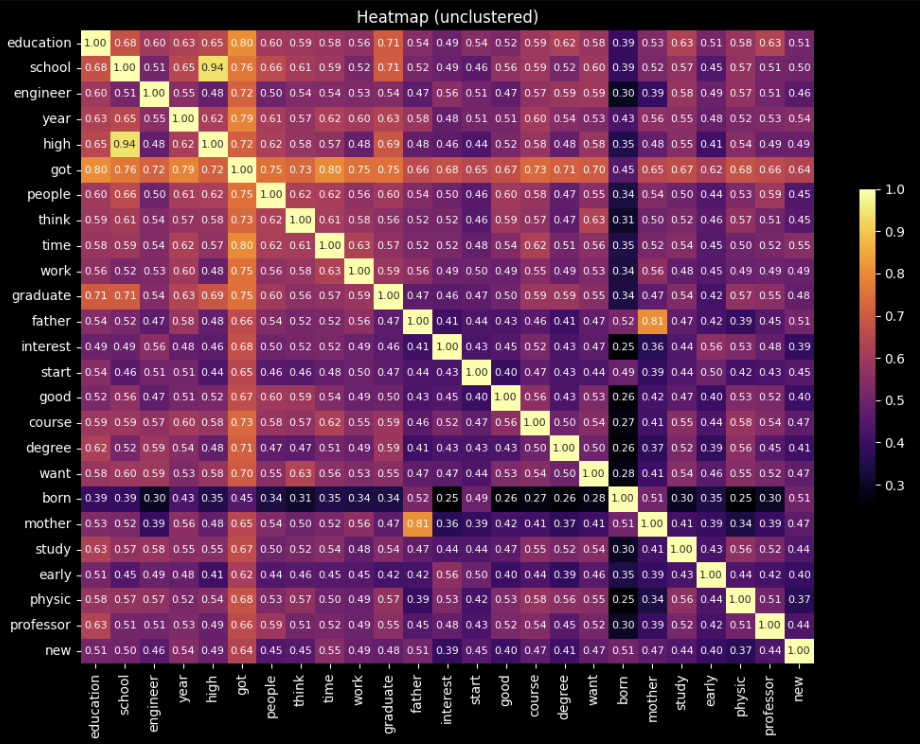}
    \captionof{figure}{Heatmap}
    \label{heatmap}
  \end{minipage}\hfill
  \begin{minipage}{0.5\textwidth}
    \centering
    \includegraphics[width=\linewidth]{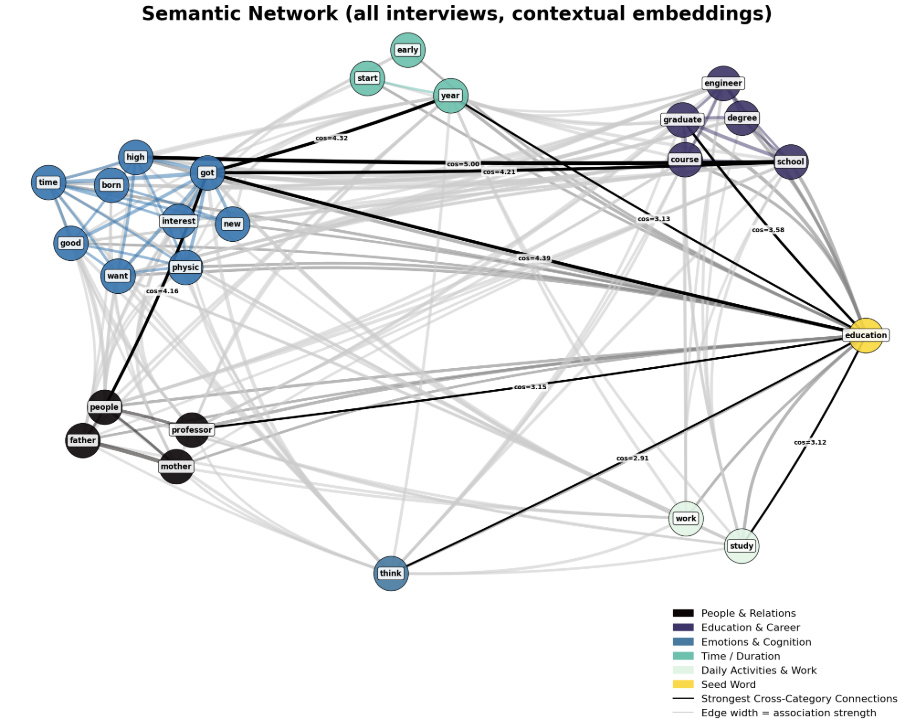}
    \captionof{figure}{Semantic Network}
    \label{semantic_network}
  \end{minipage}

\end{figure}

\begin{itemize}
    \item \textbf{Word Clouds} (\autoref{word_cloud}): Highlight the most frequent and salient terms across a dataset or within filtered subsets, with options for color coding by theme.
    
    \item \textbf{t-SNE Semantic Maps} (\autoref{tsne}): Reduce high-dimensional similarity matrices into 2D plots, emphasizing seed words for interpretability.
    
    \item \textbf{Word Heatmaps} (\autoref{heatmap}): Show how concepts or ``codes'' (meta-data used to index text, such as \#morality\_talk) relate to each other on a color-coded table with clustering options.
    \begin{itemize}
        \item \textbf{Basic Heatmap}: Clusters keywords by similarity.
        \item \textbf{Code Co-Occurrence Heatmaps}: Display the frequency with which qualitative codes appear together in the same entries.
    \end{itemize}
    
    \item \textbf{Semantic Networks} (\autoref{semantic_network}): Visualize relationships among codes or concepts as nodes and edges, with edge weights reflecting co-occurrence or similarity. Users can define custom semantic groups inductively (e.g., from heatmaps or close readings) or deductively (via theory-driven categories). Normalized cosine similarity scores (1--5) highlight the strongest links between clusters, with options for styling (color, edge thickness, clustering). Networks are typically paired with heatmaps for inductive cross-reference.
    \begin{itemize}
        \item \textbf{Heatmap + Network (Plain)}: Overlays a basic network on the heatmap.
        \item \textbf{Heatmap + Network (Colored)}: Adds colored clusters, semantic links, and optional edge styling.
    \end{itemize}
\end{itemize}

\begin{figure}[H]
    \centering
    \includegraphics[width=\textwidth]{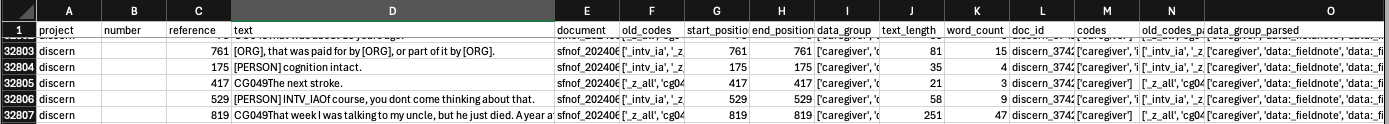}
    \caption{Example in \texttt{.csv}.}
    \label{example_csv}
\end{figure}

The examples shown below (\autoref{example_csv}) use qualitative interview data described in \cite{abramson2015beyond}, but CMAP can be applied to any properly formatted \texttt{.csv} dataset.

\begin{figure}[H]
    \centering
    \includegraphics[width=0.7\textwidth]{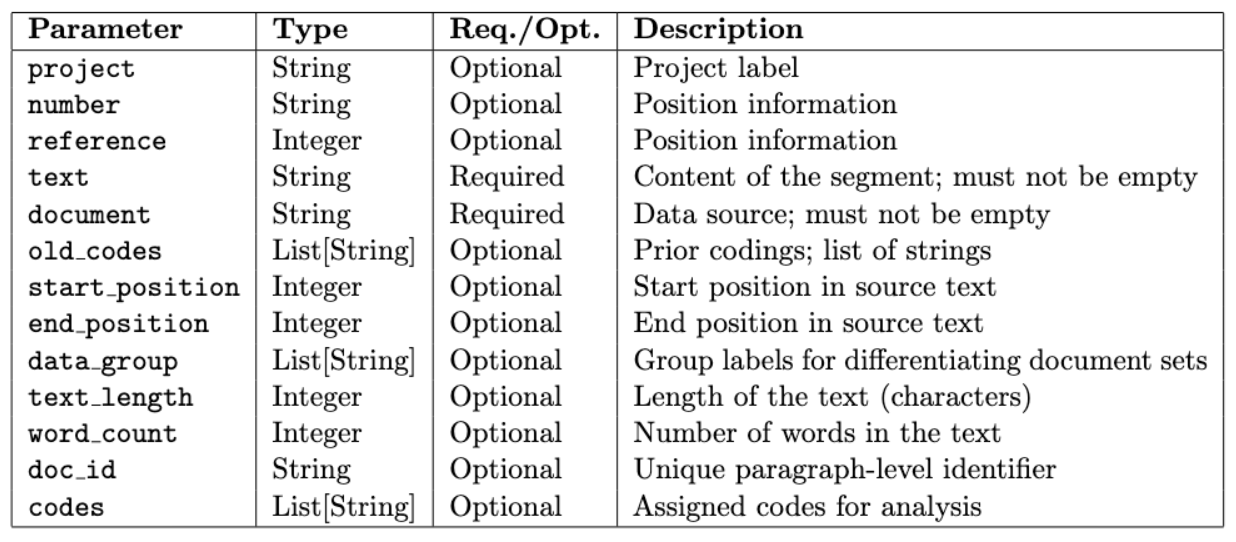}
    \caption{Parameter schema for CMAP text segments.}
    \label{schema}
\end{figure}

Any data can be used as long as it includes the required fields from the schema below (\autoref{schema}).

\begin{figure}[H]
    \centering
    \includegraphics[width=0.7\textwidth]{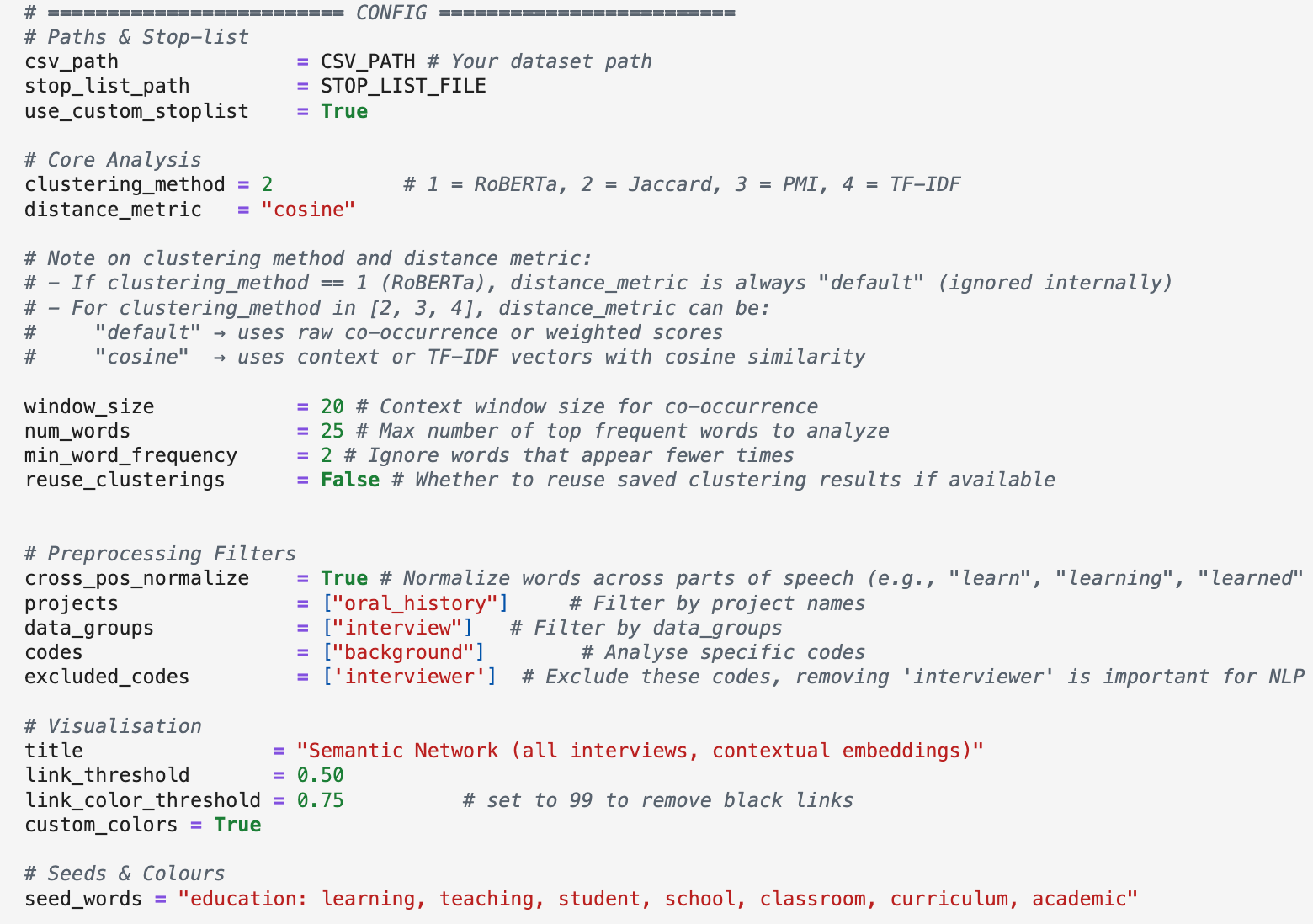}
    \caption{Configurable Execution Block.}
    \label{config}
\end{figure}

All parameters are configurable in labeled execution blocks (\autoref{config}), which determine how the visuals are produced. 
For instance, short text windows with few words are suited for syntactic analyses, while larger windows and more seeds can reveal overlapping themes. 
Users can designate colored groupings to correspond to deeper readings of text \cite{abramson2024inequality}, or use lists to compress concepts earlier in the pipeline.

\section*{Conclusion}

CMAP addresses a critical gap in open-source, scalable analytical tools for qualitative researchers, 
providing transparent statistical foundations suitable for both research publication and pedagogical applications.

\section*{Acknowledgements}

Aspects of this research were supported by the National Institute on Aging of the National Institutes of Health (NIA/NIH) award DP1AG069809 (Dohan, PI). 
Content and views are those of the authors and not of the NIH.

We thank Daniel Dohan, Zhuofan Li, Tara Prendergast, Kieran Turner, Jakira Silas, Kelsey Gonzalez, Alma Hernandez, Ignacia Arteaga, Melissa Ma, Brandi Ginn, and Zain Khemani for their feedback. 
We also acknowledge participants in \emph{“Trends in Mixed-Methods Research”}, a panel on \emph{“Computational and Mathematical Approaches to Qualitative and Quantitative Data”} organized by Laura Nelson at the American Sociological Association, 
and attendees of workshops including \emph{An Introduction to Machine Learning for Qualitative Research} and the \emph{American Sociological Association Methodology Workshop} (with Li and Dohan).

\section*{Author Contributions}

C.M.A. led project conceptualization, software architecture, software development, manuscript preparation, and test of teaching materials. Y.N. contributed equally to manuscript writing, preparation and software implementation. Both authors contributed to all aspects of the work including writing and testing code.

\bibliographystyle{plain}
\bibliography{main}

\newpage
\section*{Appendix}
\subsection*{Statistics}

\subsubsection*{RoBERTa}

We employ RoBERTa, a transformer-based language model \cite{liu2019roberta}, to obtain contextual token embeddings. 
Each paragraph in the corpus is tokenized, and subword tokens are mapped to hidden states from the final layer of the model. 
Consecutive subword tokens belonging to the same lexical unit are aggregated into word-level embeddings by averaging their hidden state vectors. 
To reduce morphological variance, each word is lemmatized.

Formally, let $x = (t_1, t_2, \dots, t_n)$ denote a sequence of tokens and $\mathbf{h}_i \in \mathbb{R}^d$ the hidden representation of token $t_i$ from the final layer of RoBERTa. 
For a word $w$ composed of tokens $\{t_{i}, \dots, t_{j}\}$, its embedding is

\[
\mathbf{v}_w = \frac{1}{j-i+1} \sum_{k=i}^{j} \mathbf{h}_k
\]

All occurrences of a word across the corpus are then averaged to form its document-level representation:

\[
\bar{\mathbf{v}}_w = \frac{1}{N_w} \sum_{m=1}^{N_w} \mathbf{v}_w^{(m)},
\]

where $N_w$ is the number of times word $w$ appears.

To identify candidate words most semantically related to the seed set $S$, we compute the cosine similarity between embeddings:

\[
\mathrm{cos}(\mathbf{u}, \mathbf{v}) = 
\frac{\mathbf{u} \cdot \mathbf{v}}{\|\mathbf{u}\| \, \|\mathbf{v}\|}
\]

For each candidate word $c$, its score is the average similarity to the seed embeddings:

\[
\mathrm{score}(c) = \frac{1}{|S|} \sum_{s \in S} 
\mathrm{cos}(\bar{\mathbf{v}}_c, \bar{\mathbf{v}}_s)
\]

The top-ranked words by $\mathrm{score}(c)$ are selected to expand the seed set, and the resulting embeddings are used to construct a cosine similarity matrix for subsequent clustering and network analysis.

\subsubsection*{Jaccard}

Jaccard similarity measures how much two sets overlap. 
A value of $1$ means the sets are identical, while $0$ means they share nothing in common:

\[
\mathrm{Jaccard}(A,B) = \frac{|A \cap B|}{|A \cup B|}
\]

\noindent\textbf{Implementation:} For each pair of words $w_i$ and $w_j$, we collect the unique context words that appear within a sliding window around them, denoted $\mathcal{C}_{w_i}$ and $\mathcal{C}_{w_j}$. 
Their Jaccard score tells us how similar the two context sets are. 
A higher score means the words tend to appear with similar neighbors, making them more closely linked in the semantic network.

\subsubsection*{PMI and PPMI}

Pointwise Mutual Information (PMI) measures how strongly two words are linked compared to what we would expect if they were independent. 
A positive PMI means the words appear together more often than chance, while a negative PMI means they appear together less often. 
To keep the measure stable and interpretable, we use Positive PMI (PPMI), which replaces all negative values with zero. 
For example, the pair \emph{New} and \emph{York} has a high PPMI because they almost always occur together, whereas \emph{New} and \emph{banana} would have a PPMI close to zero.

\[
\mathrm{PMI}(x,y) = \log_2\frac{P(x,y)}{P(x)P(y)}, \quad
\mathrm{PPMI}(x,y) = \max(0,\mathrm{PMI}(x,y))
\]

\noindent\textbf{Implementation:} We build a co-occurrence matrix by sliding a context window across the corpus. 
From these counts we estimate probabilities $p_{ij}$, $p_i$, and $p_j$, and compute

\[
e^{\mathrm{PPMI}}_{w_i,j} = \max\left(0, \log_2 \frac{p_{ij}}{\max(\varepsilon,p_i)\max(\varepsilon,p_j)} \right)
\]

Here $p_{ij}$ is the probability that anchor $w_i$ and context word $c_j$ co-occur, and $\varepsilon$ (e.g., $10^{-10}$) prevents division by zero. 
We then apply cosine similarity to the resulting PPMI vectors to compare words in the semantic network.

\subsubsection*{TF--IDF}

Term Frequency--Inverse Document Frequency (TF--IDF) assigns higher weight to a term if it is frequent in a given context but relatively rare across the entire corpus. 
This makes it useful for identifying words that are especially informative, rather than just common.

\[
\mathrm{tfidf}(t,d,\mathcal{D}) = \mathrm{tf}(t,d) \cdot \log \frac{|\mathcal{D}|}{\mathrm{df}(t)}
\]

where $\mathrm{tf}(t,d)$ is the frequency of term $t$ in document $d$, and $\mathrm{df}(t)$ is the number of documents containing $t$ in the corpus $\mathcal{D}$.

\noindent\textbf{Implementation:} For anchor $w_i$ and context $c_k$ with raw count $v_{w_i,k}$, we weight each context by its TF--IDF score:

\[
e^{\mathrm{T}}_{w_i,k} = v_{w_i,k} \cdot \mathrm{tfidf}(c_k)
\]

Cosine similarity between rows of $e^{\mathrm{T}}$ gives an anchor-to-anchor similarity matrix, showing how strongly two words are connected through their distinctive contexts. 
For example, \emph{doctor} and \emph{hospital} may yield a high similarity score because they share informative context words, while \emph{doctor} and \emph{banana} will score low.

\subsubsection*{Raw Context--Count Vectors}

The simplest way to represent a word is to count how often other words appear near it. 
For each target word $w_i$, we slide a fixed window of size $w$ across the corpus. 
Every time $w_i$ occurs, we look at the surrounding context words in that window (including $w_i$ itself) and add one to their counts. 
This gives a vector $\vec{v}_{w_i}$ where each entry $v_{w_i,k}$ records how often context word $c_k$ appears near $w_i$.

\[
v_{w_i,k} = \sum_{\text{sent}\in\mathcal{D}} \sum_{\substack{p:\,\text{sent}[p]=w_i}} \sum_{q=\max(0,p-w)}^{\min(|\text{sent}|-1,p+w)} \mathbf{1}\{\text{sent}[q]=c_k\}, \quad 
\vec{v}_{w_i}=(v_{w_i,1},\dots,v_{w_i,n})
\]

After building these vectors, we compute cosine similarity between them to measure how similar two words’ contexts are. 
For example, if \emph{doctor} and \emph{nurse} often appear near similar words (\emph{hospital, patient, care}), their vectors will be close, and cosine similarity will assign them a high score.

\subsection*{Distance Metric}

Given $E^{\phi} \in \mathbb{R}^{m \times n}$ with rows $(\vec{e}^{\,\phi}_{w_i})^{T}$, we define the similarity matrix as

\[
S^{\phi}_{ij} = \cos(\vec{e}^{\,\phi}_{w_i}, \vec{e}^{\,\phi}_{w_j}) = 
\frac{\vec{e}^{\,\phi}_{w_i} \cdot \vec{e}^{\,\phi}_{w_j}}{\|\vec{e}^{\,\phi}_{w_i}\| \, \|\vec{e}^{\,\phi}_{w_j}\|}
\]

Cosine similarity measures how close two word vectors are in direction, regardless of their magnitude. 
For two embeddings $\vec{e}^{\,\phi}_{w_i}$ and $\vec{e}^{\,\phi}_{w_j}$, it is defined as the cosine of the angle between them. 
Values near $1$ indicate strong semantic similarity, while values near $0$ or negative suggest weak or opposite meaning. 

For example, the vectors for \emph{doctor} and \emph{nurse} would yield a high cosine similarity, reflecting their related meanings, 
whereas \emph{doctor} and \emph{banana} would yield a value close to $0$. 
This makes cosine similarity a simple and effective tool for comparing words in our semantic network analysis.

\subsection*{Other Resources}

\subsubsection*{Workflow Steps Example (End-to-End)}

The workflow for analyzing text as data is iterative. 
This synthesized workflow integrates pragmatic qualitative steps \cite{abramson2025pragmatic, li2025ethnography} 
with frameworks established in computational social science (CSS) \cite{grimmer2022text}.

\begin{itemize}
    \item \textbf{Define Question / Theory:} 
    Specify the research question or Quantity of Interest (QoI). 
    Work may begin inductively \cite{nelson2020computational} or deductively \cite{grimmer2022text}.

    \item \textbf{Aggregation (Building the Corpus):} 
    Define the population, sampling frame, and document units. 
    Data sources can include transcribed interviews, ethnographic fieldnotes, historical documents, webscraped data, policy documents, administrative text, or open-ended survey responses. 
    Record provenance and metadata.

    \begin{quote}
        \textit{Python Tools:} \texttt{pandas} for manifests; \texttt{requests} + \texttt{beautifulsoup4} for web scraping; or API clients. 
        Store as JSONL/CSV + raw text. Export from QDA software or integrate text into a DataFrame.
    \end{quote}

    \item \textbf{Digitization and Processing (Data Wrangling):}

    \textit{Digitization (OCR \& QA):} Convert PDFs or scans and perform manual Quality Assurance (QA). 
    Choose digitization methods that preserve meaningful structure (e.g., speaker turns, page breaks) for citation integrity.

    \textit{Processing:} Clean and format text into machine-readable and tabular formats (see Schema below). 
    Data can be imported from QDA software or read directly from \texttt{.txt} (UTF-8) files. Tokenize, segment, and normalize.

    \begin{quote}
        \textit{Python Tools:} \texttt{pytesseract} (OCR); \texttt{spaCy} (normalization / tokenization).
    \end{quote}

    \item \textbf{Representation:} 
    Transform text into formats suitable for computational analysis. 
    Choose representations (e.g., Bag-of-Words / TF--IDF, dictionaries, embeddings) to fit the QoI. 
    This often involves visualizing patterns combined with close readings.

    \begin{quote}
        \textit{Python Tools:} \texttt{scikit-learn} vectorizers (DTM / TF--IDF); Hugging Face Transformers (embeddings).
    \end{quote}

    \item \textbf{Annotating and Linking:}

    \textit{Annotating:} Build a human system for indexing data. 
    Utilize a hybrid approach---combining automation (lists, machine learning) and human coding depending on scope and complexity \cite{abramson2025pragmatic}. 
    This involves tradeoffs between accuracy, efficiency, and discovery of insights. 
    Entity tagging (persons / organizations / places) can be performed with \texttt{spaCy} NER.

    \textit{Linking:} Join texts to variables in a dataframe (e.g., site, time, treatment, demographics) for comparison and modeling. 
    If using qualitative software or purposeful file naming, this step can often be completed with minimal work \cite{li2025ethnography}.

    \begin{quote}
        \textit{Python Tools:} \texttt{spaCy} (NER); \texttt{pandas} (linking).
    \end{quote}

    \item \textbf{Analysis, Modeling, \& Visualization:}

    \textit{Descriptions \& Visualization:} 
    LDA topic modeling with human validation (“Reading Tea Leaves”); 
    word embeddings for schema detection combined with in-depth narrative \cite{abramson2024inequality}. 
    Use visualization tools (e.g., CMAP) to explore and compare patterns \cite{abramson2015beyond}.

    \begin{quote}
        \textit{Modeling:} Supervised coding or stance detection using \texttt{scikit-learn} baselines and Transformer-based models (BERT-class). 
        Report metrics, calibration, and error analysis. Combine unsupervised exploration (topics, clusters) with supervised measurement or prediction. \\
        \textit{Deep Reading \& Interpretation:} Return to exemplar passages to contextualize model patterns, examine disconfirming cases, and refine explanations while noting contextual limits.
    \end{quote}

    \item \textbf{Dissemination and Archiving:}
    Produce reproducible Jupyter Notebooks (see workshop repository), CMAP visualizations, codebooks, and curated quotations. 
    Pair patterns with passages in presentation. 
    Archive code and data where permissible, ensuring de-identification and ethical documentation.
\end{itemize}

\subsubsection*{Data Schema Example (CMAP)}

For structured analysis and visualization (e.g., using the CMAP toolkit), data should be organized into a consistent tabular format (e.g., CSV or DataFrame). 
Below is an example schema:

\begin{verbatim}
# Updated schema with Python typing
schema = {
    "project": str,          # List project
    "number": str,           # Position information
    "reference": int,        # Position information
    "text": str,             # Content, critical field: must not be empty
    "document": str,         # Data source, critical field: must not be empty
    "old_codes": list[str],  # Optional: codings, must be a list of strings
    "start_position": int,   # Position information
    "end_position": int,     # Position information
    "data_group": list[str], # Optional: to differentiate document sets
    "text_length": int,      # Optional: NLP info
    "word_count": int,       # Optional: NLP info
    "doc_id": str,           # Optional: unique paragraph-level identifier
    "codes": list[str]       # Critical for analyses with codes
}
\end{verbatim}

\subsubsection*{Modes of Combining Computation and Qualitative Analysis}

A key consideration is how—or whether—to integrate computational tools into the analytical workflow. 
Researchers adopt different modes based on project needs, data sensitivity, and analytical goals \cite{abramson2025pragmatic}.

\begin{itemize}
    \item \textbf{Streamline (Organizational):} 
    Using computational tools to manage logistics of research---organizing manifests, facilitating de-identification, managing quotes, automating basic indexing, and tracking team progress---even when core coding remains manual.

    \item \textbf{Scaling-Up (Efficiency / Size):} 
    For large, longitudinal, or multi-site corpora, machine learning (e.g., supervised classification) can assist human coding. 
    These approaches require high-quality labeled data and rigorous validation through hybrid human–machine workflows.

    \item \textbf{Hybrid (Iterative Refinement and Mixed Methods):} 
    Combining human analysis with computational methods to answer different types of questions or refine understanding. 
    Often used in computational ethnography or historical analysis, this approach leverages computational patterns (e.g., clustering, visualization) to guide iterative reading and comparison \cite{abramson2025pragmatic}.

    \item \textbf{Discovery (Pattern Finding):} 
    Utilizing unsupervised methods (e.g., topic modeling, clustering, visualization) to identify latent patterns or typologies that guide inductive theory development \cite{nelson2020computational}.

    \item \textbf{Minimal / No Computation (The “Sociology of Computation”):} 
    Choosing not to automate analysis when ethical or interpretive considerations dominate. 
    Documenting the rationale for this choice enhances transparency and reflexivity \cite{abramson2025pragmatic}.
\end{itemize}

\subsection*{Related Software Resources}

\noindent \textbf{Li, Zhuofan and Corey M. Abramson.} 2022. 
\textit{An Introduction to Machine Learning for Qualitative Research.} 
Jupyter Notebooks (Python). American Sociological Association Methodology Workshop. 
\href{https://github.com/lizhuofan95/ASA2022_Workshop}{GitHub Repository}.

\medskip

\noindent \textbf{Nelson, Laura K.} 2020. 
“Computational Grounded Theory: A Methodological Framework.” 
\textit{Sociological Methods \& Research} 49(1):3–42. 
\href{https://journals.sagepub.com/doi/10.1177/0049124117729703}{Article} \,|\, 
\href{https://www.lauraknelson.com/}{Homepage}.

\medskip

\noindent \textbf{Commercial Qualitative Data Software} 
(limited scalability for large datasets, lacks advanced CSS/statistical methods, and/or requires cloud computing):

\begin{itemize}
    \item ATLAS.ti Scientific Software Development GmbH. 2023. \textit{ATLAS.ti Mac} (version 23.2.1). \url{https://atlasti.com}
    \item Dedoose Version 9.0.107. 2023. Los Angeles, CA: SocioCultural Research Consultants, LLC. \url{https://www.dedoose.com}
    \item Lumivero. 2023. \textit{NVivo} (Version 14). \url{https://www.lumivero.com}
\end{itemize}

\subsection*{Representative Scientific Applications}

\subsubsection*{Peer-Reviewed Articles}

\begin{itemize}
    \item Abramson, Corey M., Tara Prendergast, Zhuofan Li, and Martín Sánchez-Jankowski. 2024. 
    “Inequality in the Origins and Experiences of Pain: What ‘Big (Qualitative) Data’ Reveal About Social Suffering in the United States.” 
    \textit{Russell Sage Foundation Journal of the Social Sciences} 10(5):34–65. 
    \href{https://www.rsfjournal.org/content/rsfjss/10/5/34.full.pdf}{Link}.
    
    \item Arteaga, Ignacia, Alma Hernández de Jesús, Brandi Ginn, Corey M. Abramson, and Daniel Dohan. 2025. 
    “Understanding How Social Context Shapes Decisions to Seek Institutional Care: A Qualitative Study of Experiences of Progressive Cognitive Decline Among Latinx Families.” 
    \textit{The Gerontologist} gnaf207. 
    \href{https://doi.org/10.1093/geront/gnaf207}{Link}.
    
    \item Li, Zhuofan and Corey M. Abramson. 2025. 
    “Ethnography and Machine Learning: Synergies and Applications.” 
    In \textit{Oxford Handbook of the Sociology of Machine Learning}, edited by [editors]. Oxford University Press. 
    \href{https://arxiv.org/abs/2412.06087}{Preprint}.
    
    \item Abramson, Corey M., Zhuofan Li, and Tara Prendergast. Expected 2026. 
    “Qualitative Research in an Era of AI: A Pragmatic Approach to Data Analysis, Workflow, and Computation.” 
    \textit{Annual Review of Sociology.} 
    \href{https://arxiv.org/pdf/2509.12503}{Preprint available}.
\end{itemize}

\subsubsection*{Conference Presentations (2024–2025)}

\begin{itemize}
    \item Abramson, Corey M., Kieran Turner, Ignacia Arteaga, Alma Hernández de Jesús, Brandi Ginn, Yuhan Nian, and Daniel Dohan. 2025. 
    “Pragmatic Sensemaking: Semantic Maps of Dementia Narratives.” 
    \textit{ARS’25: Tenth International Workshop on Social Network Analysis.} Naples, Italy.

    \item Abramson, Corey M., Kieran Turner, Ignacia Arteaga, Alma Hernández de Jesús, Brandi Ginn, Yuhan Nian, and Daniel Dohan. 2025. 
    “Pragmatic Sensemaking: Mapping the Cultural Work of Living with Dementia.” 
    \textit{American Sociological Association Annual Meeting.} Chicago, IL.

    \item Abramson, Corey M., Zhuofan Li, and Tara Prendergast. 2024. 
    “Qualitative Sociology in a Computational Era: Classic Issues, Emerging Trends, and New Possibilities.” 
    \textit{American Sociological Association Annual Meeting.} Montreal, Canada.
\end{itemize}
\end{document}